\documentclass[%
 reprint,
 amsmath,amssymb,
 aps,
pra,
]{revtex4-2}

\usepackage{graphicx}%
\usepackage{dcolumn}%
\usepackage{bm}%

\usepackage{braket}
\usepackage{amsmath}
\usepackage{subfigure}
\usepackage{enumitem}

\usepackage[normalem]{ulem}  %

\newcommand{\st}{\mathop{\mathrm{s.t.}}}
\newcommand{\mat}[1]{\mathbf{#1}}
\newcommand{\vect}[1]{\boldsymbol{#1}}
\def\R{\mathbb{R}}

\newcommand{\Gate}[1]{\textsc{#1}}

\newcommand{\ygate}{\Gate{y}}
\newcommand{\xgate}{\Gate{x}}

\newcommand{\cnotgate}{\Gate{cnot}}

\makeatletter
\newcommand{\vast}{\bBigg@{3}}
\newcommand{\Vast}{\bBigg@{4}}
\makeatother

\usepackage{hyperref}%

\usepackage{xcolor}
\usepackage{lineno}
\newcommand{\XY}{\xgate\ygate}

\begin{document}

\title{Alignment between Initial State and Mixer Improves QAOA Performance \\ for Constrained Optimization} %

\author{Zichang He}
\email{zichang.he@jpmorgan.com}
\affiliation{
 Global Technology Applied Research, JPMorgan Chase, New York, NY 10017 USA
}
\author{Ruslan Shaydulin}
\affiliation{
 Global Technology Applied Research, JPMorgan Chase, New York, NY 10017 USA
}
\author{Shouvanik Chakrabarti}
\affiliation{
 Global Technology Applied Research, JPMorgan Chase, New York, NY 10017 USA
}
\author{Dylan Herman}
\affiliation{
 Global Technology Applied Research, JPMorgan Chase, New York, NY 10017 USA
}
\author{Changhao Li}
\affiliation{
 Global Technology Applied Research, JPMorgan Chase, New York, NY 10017 USA
}
\author{Yue Sun}
\affiliation{
 Global Technology Applied Research, JPMorgan Chase, New York, NY 10017 USA
}
\author{Marco Pistoia}
\affiliation{
 Global Technology Applied Research, JPMorgan Chase, New York, NY 10017 USA
}

\begin{abstract}
Quantum alternating operator ansatz (QAOA) has a strong connection to the adiabatic algorithm, which it can approximate with sufficient depth. However, it is unclear to what extent the lessons from the adiabatic regime apply to QAOA as executed in practice with small to moderate depth. In this paper, we demonstrate that the intuition from the adiabatic algorithm applies to the task of choosing the QAOA initial state. Specifically, we observe that the best performance is obtained when the initial state of QAOA is set to be the ground state of the mixing Hamiltonian, as required by the adiabatic algorithm. We provide numerical evidence using the examples of constrained portfolio optimization problems with both low ($p\leq 3$) and high ($p = 100$) QAOA depth. Additionally, we successfully apply QAOA with $\XY$ mixer to portfolio optimization on a trapped-ion quantum processor using 32 qubits and discuss our findings in near-term experiments.

\end{abstract}

\maketitle

\section{Introduction}\label{sec:intro} 

Combinatorial optimization is one of the most promising applications of quantum computers due to its broad applicability in science and industry and the availability of promising quantum algorithms with the potential for speedups over the classical state-of-the-art~\cite{shaydulin2023evidence,2208.06909}.
A leading quantum algorithm for combinatorial optimization is the quantum approximate optimization algorithm~\cite{farhi2014quantum,hogg2000quantum} 
and its generalization, quantum alternating operator ansatz (QAOA)~\cite{hadfield2019quantum}. QAOA solves the optimization problem by preparing a parameterized quantum state using a quantum circuit consisting of layers of phase and mixing (mixer) operators applied in alternation, with parameters optimized to extremize a chosen measure of solution quality. 
QAOA has promising applications in optimization~\cite{wang2020xy,shaydulin2021qaoakit,harrigan2021quantum,tomesh2022quantum,saleem2023approaches,Pelofske2023,2301.11292}, finance~\cite{herman2022survey,herman2022portfolio} and machine learning~\cite{benedetti2019parameterized,Niroula2022,otterbach2017unsupervised}, and has been adapted to be applicable to quantum chemistry~\cite{kremenetski2021quantum}.
Among the numerous QAOA variants that have been introduced, our focus is on the widely-studied local Hamiltonian-based QAOA (LH-QAOA)~\cite{hadfield2019quantum}. In this variant, the phase and mixing operators correspond to the time evolution under phase Hamiltonian $\mat{H}_P$ and mixing Hamiltonian $\mat{H}_M$, with $\mat{H}_M$ being the sum of polynomially many local terms. Note that $\mat{H}_P$ does not have to be local.
In the remainder of the paper, we use QAOA to refer to LH-QAOA. 

QAOA has an important connection to adiabatic quantum algorithm (AQA)~\cite{farhi2000quantum,Albash2018}. AQA prepares ground states of Hamiltonians by performing a slow interpolation between an easy-to-prepare ground state of some simple Hamiltonian and the ground state of the target Hamiltonian. The speed of interpolation is governed by the minimum spectral gap of the instantaneous system Hamiltonian during the evolution. AQA can be applied to optimization problems by choosing an appropriately constructed diagonal Hamiltonian as the target. If the alternating operators in QAOA are chosen to be time evolution with the target and a simple Hamiltonian (e.g., the commonly used transverse field Hamiltonian), and the initial state is set to be the ground state of the simple Hamiltonian, QAOA can approximate the AQA evolution with an approximation error that depends on the number of the alternating layers (QAOA depth).

While the connection between AQA and QAOA is simple and well-known, QAOA is typically used with parameters that are different from the AQA schedule and with a depth that is too small to approximate AQA meaningfully. This creates ambiguity regarding the extent to which the QAOA mechanism is related to AQA, as well as how to leverage the techniques for boosting AQA performance in the QAOA setting. In this paper, we show that in one important aspect, the lessons of the adiabatic algorithm indeed apply clearly. Namely, we show that QAOA performance is improved if the ground state of $\mat{H}_M$ and the initial state are aligned. Specifically, we show that QAOA gives better performance when the initial state matches the ground state of $\mat{H}_M$ compared to other setups. This choice of initial state and $\mat{H}_M$ also aligns with that in the AQA. We refer to this setup as initial-mixer alignment or alignment for short.

We note that in some cases the mixer as well as the ground state of $\mat{H}_M$ are difficult to implement on the quantum hardware. Therefore is may be desirable to use alternative initial states and mixers that are not well-aligned but are easy to implement. Moreover, previous studies have found that the performance of QAOA may be improved by carefully preparing a `warm-start' initial state different from the ground state of $\mat{H}_M$ ~\cite{egger2021warm,tate2020bridging}. In general, it may be hard to modify the mixer to make sure that the warm-start initial state is exactly the ground state of the mixer. These examples motivate the current study of misaligned combinations of mixer and initial state.

In this work, we study QAOA with Hamming-weight-preserving $\XY$ mixers~\cite{wang2020xy}, where the mixing operator is a time evolution governed by Heisenberg $\XY$ models~\cite{lieb1961two}. This variant of QAOA is of particular interest as evidence suggests that it has the potential to provide exponential speedup over unstructured search on certain problems~\cite{2202.00648}. We choose the various $\XY$ models as the constraint-preserving mixing Hamiltonian in QAOA: ring-$\XY$, complete-$\XY$, and several $\XY$ models with arbitrary connectivity. We apply QAOA with $\XY$ mixers to the portfolio optimization problem with an equality constraint on the portfolio size, which corresponds to a constraint on the Hamming weight of the binary string. This is a well-studied toy financial problem, which is commonly considered as a benchmark for quantum optimization heuristics~\cite{herman2022portfolio,brandhofer2023benchmarking,slate2021quantum,hodson2019portfolio,Hao2022,baker2022wasserstein}. Such heuristic algorithms often aim at approximately solving the portfolio optimization problem with a goal of maximizing the approximation ratio. We will follow the same convention and use approximation ratio as the primary metric for evaluating the performance of QAOA. To quantify the impact of alignment on QAOA performance, we design two sets of numerical experiments. First, we compare the performance of various initial-mixer pairs. Here, we isolate the impact of alignment by considering the exact implementation of the mixer. Second, we fix the initial state as the ground state of $\mat{H}_M$ and implement the $\XY$ mixer with various fidelities by varying the step of Trotter approximation. This setup highlights the practical considerations when implementing complex mixers with non-commuting terms in $\mat{H}_M$.

\begin{figure}
    \centering
    \includegraphics[width = 3.3in]{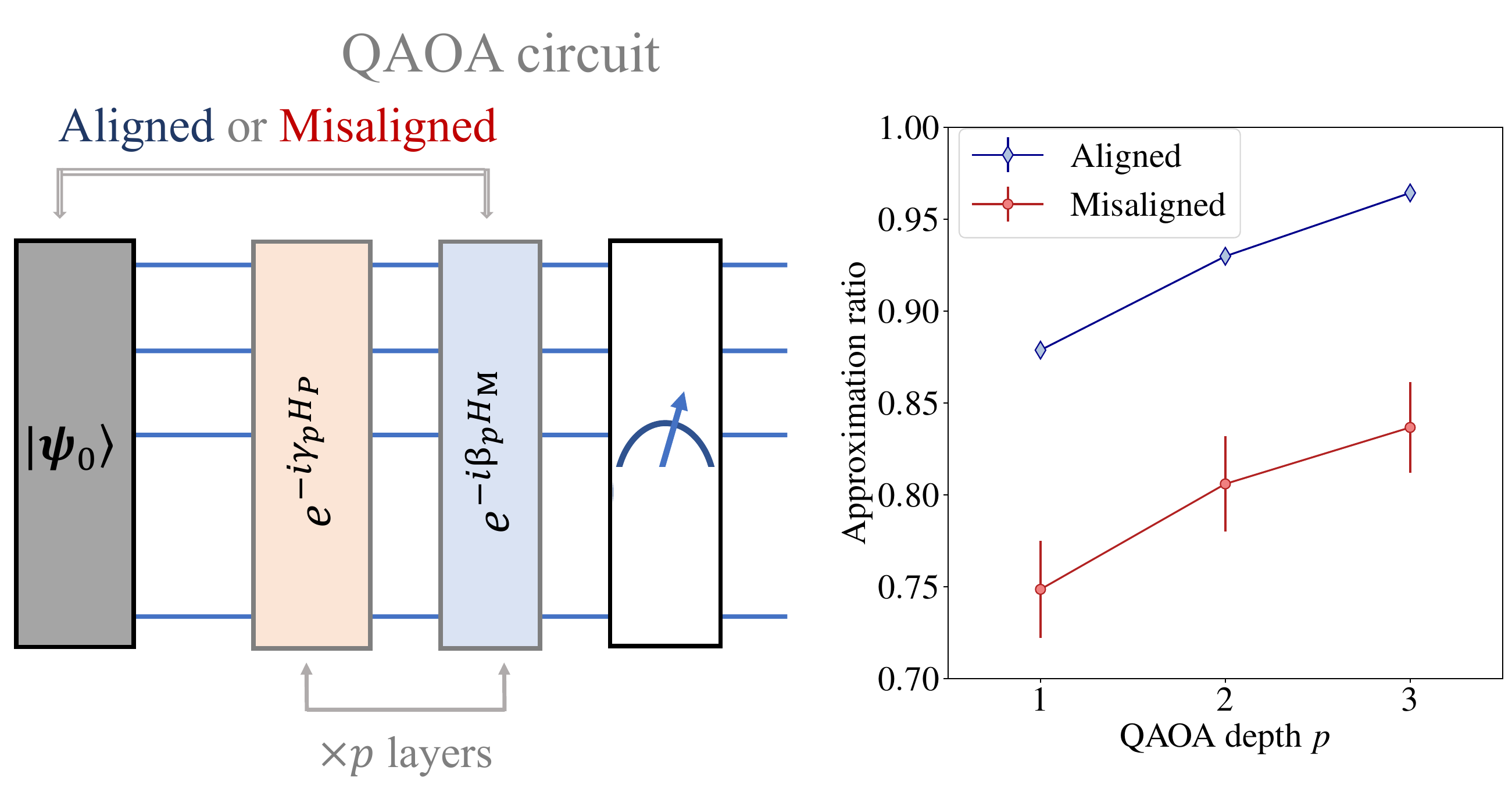}
    \caption{An overview of the results. We show that the QAOA performance depends on the alignment between the initial state $\ket{\psi_0}$ of QAOA and the ground state of the mixing Hamiltonian $\mat{H}_M$. Right: the approximation ratios (ARs) obtained by QAOA applied to constrained portfolio optimization with $N=6$ assets and Hamming weight constraint $K=3$ with the complete-$\XY$ mixer and the initial state set to be the ground state of complete-$\XY$ (``Aligned'') and ring-$\XY$ (``Misaligned'') mixing Hamiltonian. The error bars represent the standard error of the mean approximation ratio estimated from $10$ problem instances of portfolio optimization.} %
    \label{fig:overview}
\end{figure}

The main conclusion of the paper is illustrated in Fig.~\ref{fig:overview}. Our results show that, in most cases, the alignment boosts the QAOA performance. The only exception is when the mixing $\XY$ Hamiltonian is relatively simple, e.g., a chain or ring. In these cases, QAOA performance is more robust under Trotter approximation error. However, given a high Trotter error, a more accurate Trotter approximation still improves the QAOA performance. Across all simulations, we observe a consistent trend in QAOA with both low and high depth. The first set of results clearly shows the alignment effect without considering the circuit implementation. 

While the improvement in performance from alignment is consistent across the many settings we consider, its absolute effect is relatively small. Therefore, when executing on noisy intermediate-scale quantum era (NISQ) devices, enhancing the alignment at the cost of increased circuit depth is unlikely to improve the results significantly. To illustrate this observation, we apply QAOA with ring-$\XY$ mixer to portfolio optimization using all $32$ qubits of the Quantinuum H2-1 trapped-ion processor. We observe that even the step-1 Trotter approximation of the ring-$\XY$ mixer gives a high-quality solution on hardware, and further improvements in alignment do not significantly increase the solution quality on hardware. This contrasts with the noiseless case, where a more accurate Trotter approximation results in better performance.

To the best of our knowledge, this is also the first study of the impact of mixer Trotterization on QAOA performance. Recent works have developed various techniques to improve the performance of QAOA. Specifically, non-standard initial states have been used, such as the `warm-start' initial state constructed using a solution produced by a classical solver~\cite{egger2021warm,tate2020bridging} abd the randomly sampled computational basis states with a given Hamming weight for QAOA with Hamming-weight preserving mixers~\cite{cook2020quantum}. In addition, alternative ans{\"a}tze have also been proposed, such as initial-state-dependent custom mixers for warm-started QAOA~\cite{tate2021classically}. Here, we do not aim to propose an optimal ansatz with minimal depth but try to systematically demonstrate the mechanism that the alignment effect from the adiabatic theorem applies to QAOA in the low-depth regime. 
Therefore, our study will not include techniques in which the initial state and the ground state of the mixing Hamiltonian are purposefully not aligned, such as the ones mentioned above.
Beyond the investigation of the QAOA mechanism, we also discuss the techniques we used to improve the convergence of local QAOA parameter optimizers, which may be of independent interest. We note that our results are expected to apply broadly beyond the particular problem considered and may be particularly impactful in applications where the target ground state must be prepared with high fidelity, such as in quantum chemistry~\cite{kremenetski2021quantum}. 

\section{Results}\label{sec:results}
\subsection{Background}\label{sec:background}
In this section, we will briefly review the relevant technical background around the portfolio optimization problem, the quantum alternating operator ansatz (QAOA) and the adiabatic quantum algorithm (AQA). We will also discuss parameter optimization for QAOA and the connection between QAOA and AQA.

Portfolio optimization problem.
We focus on the mean-variance portfolio optimization problem~\cite{markowitz_port} with objective $f$ given by
\begin{equation}\label{eq:portfolio_opt}
\begin{aligned}
\min_{\mat{x} \in \{0,1\}^N} f(\mat{x}) & = q \mat{x}^T \mat{W} \mat{x} - \vect{\mu}^T \mat{x}, \\
\st \quad & \mat{1}^T \mat{x} = K,    
\end{aligned}
\end{equation}
where $\mat{x}\in \{0,1\}^N$ denotes a vector of binary decision variables indicating whether a given asset is included in ($1$) or excluded from ($0$) the portfolio, 
$\vect{\mu} \in \R^{N}$ denotes the vector of expected returns for the assets, $\mat{W} \in \R^{N \times N}$ is the covariance matrix between $N$ assets and $q>0$ is a risk factor to balance the importance of risk and return in the objective. The equality constraint corresponds to a fixed budget requiring the manager to pick exactly $K$ assets. %
This equality constraint is also called a Hamming-weight-preserving constraint since it restricts the Hamming weight of $\mat{x}$ to a constant $K$. 

Quantum alternating operator ansatz (QAOA).
In order to apply QAOA for solving~\eqref{eq:portfolio_opt}, we must define the Hamiltonians in the two alternating operators: $\mat{H}_P$ encoding the classical objective function and $\mat{H}_M$ mixing the probability amplitudes while preserving the constraints. 

To encode the objective function \eqref{eq:portfolio_opt}, we construct a diagonal Hamiltonian $\mat{H}_P=\text{diag}(f(\mat{x}))$ by mapping each binary variable $x_i$ to a quantum spin using $x_i \to ({\mat{I}-\mat{Z}_i})/{2}$, giving 
\begin{equation}\label{eq:problem_Hamitonian}
    \mat{H}_P =\frac{1}{2}q \sum_{i < j} W_{ij} \mat{Z}_i \mat{Z}_j - \frac{1}{2}\sum_i (q\sum_{j} W_{ij} - \mu_i)\mat{Z}_i + c,
\end{equation}
where ${c = \frac{1}{2} \sum_{i} (q\sum_{j=i} W_{ij} - \mu_i})$ is a constant. 
We denote the time evolution under $\mat{H}_{P}$ given by $e^{-i \beta \mat{H}_P}$ as the phase operator.

In order to enforce the Hamming weight constraint on the quantum state, we follow Refs.~\cite{hadfield2019quantum,wang2020xy} and use the Heisenberg $\XY$ model as the mixing Hamiltonian
\begin{equation}
    \mat{H}_M:=\mat{H}_{S}^{\XY} = \sum_{(i,j)\in S} \mat{X}_{i}\mat{X}_{j} + \mat{Y}_{i}\mat{Y}_{j},
\end{equation}
where $S$ is a set of index pairs describing the interaction among qubits, $\mat{X}$ and $\mat{Y}$ are the Pauli matrices. We denote the time evolution under $\mat{H}_{S}^{\XY}$ given by $e^{-i \beta \mat{H}_S^{\XY}}$ as the $\XY$ mixer. The performance and the implementation difficulty of an $\XY$ mixer depend on the choice of the connectivity defined by $S$. Two commonly used $\XY$ models are \emph{ring-$\XY$} and \emph{complete-$\XY$}. Specifically, the ring-$\XY$ model includes one-dimensional nearest-neighbor interactions with a periodic boundary condition, i.e.,
$$ 
S_{\rm ring} = \Big\{ (i,j)~\vert~j = (i + 1) \mod N; i \in [N] \Big\}.
$$
On the other hand, the complete-$\XY$ model contains interactions between all pairs of qubits, i.e.,
$$ S_{\rm complete}=\Big\{ 
(i,j)~\vert~i < j; i,j \in [N]
\Big\}.$$
It is easy to see that the evolution with the $\XY$ mixers preserves the Hamming weight.
In other words, if we start from a superposition of states of Hamming weight $K$, the measurement outcomes of the final state are also guaranteed to have Hamming weight $K$. 

For a given pair of $\mat{H}_P$ and $\mat{H}_M$, QAOA with depth $p$ consists of the following three steps. First, QAOA prepares a feasible initial state $\ket{\vect{\psi}_0}$. Then the phase operator and the mixer
are applied $p$ times to obtain the state 
\begin{equation}
   \ket{\vect{\psi}(\vect{\gamma},\vect{\beta})} = e^{-i \beta_p \mat{H}_M} e^{-i \gamma_p \mat{H}_P} \ldots e^{-i \beta_1 \mat{H}_M} e^{-i \gamma_1 \mat{H}_P} \ket{\vect{\psi}_0}, 
\end{equation}
where $\vect{\gamma}$ and $\vect{\beta}$ are vectors of free parameters obtained using some classical procedure.
Finally, the state $\ket{\vect{\psi}(\vect{\gamma},\vect{\beta})}$ is measured in the computational basis to obtain solutions to the original problem.

QAOA is typically used as a hybrid quantum-classical algorithm wherein a classical optimizer is used to optimize the parameters $\vect{\gamma}$ and $\vect{\beta}$
to minimize the energy of $\mat{H}_P$. We denote it as $\emph{unrestricted}$ optimization:
\begin{equation}\label{eq:finite_optimize}
    \min_{\vect{\gamma},\vect{\beta}} \bra{\vect{\psi}(\vect{\gamma},\vect{\beta})} \mat{H}_P \ket{\vect{\psi}(\vect{\gamma},\vect{\beta})}.
\end{equation}
Usually, the parameter optimization is nontrivial since the energy landscape is known to contain many local optima. Many advanced methods have been developed for QAOA training~\cite{liu2022layer,shaydulin2022parameter,yao2020policy,he2023distributionally}. 
We will discuss our techniques for accelerating the parameter optimization in the method section.

When $p$ is large, parameter optimization becomes hard. Restricting the QAOA parameters can allow faster parameter optimization for large $p$. For example, it has been shown that good solution quality can be achieved with reduced optimization complexity by using a linear ramp schedule for the QAOA parameters given by~\cite{kremenetski2021quantum,hogg2000quantum,zhou2020quantum,sack2021quantum,shaydulin2021classical}
\begin{equation}
    \gamma(l) = \Delta l, \quad \beta(l) = \Delta (1 - l), 
\end{equation}
where $\Delta$ is a constant and $l \in (0,1)$. 
In QAOA with depth $p$, the linear schedule may be applied with the QAOA parameters for each layer set as follows:
\begin{equation}\label{eq:phase_diagram_schedule}
\begin{aligned}
    \gamma_i & = \gamma(l_i), \beta_i  = \beta(l_i) \quad {\rm with} \\
    l_i = &\frac{i}{p+1}, \quad \forall i=1,2,\ldots,p.
\end{aligned}
\end{equation}
With $\Delta>0$ and $p \to \infty$, the QAOA approaches the adiabatic limit~\cite{hogg2003adiabatic}. In this case, if the initial state is the ground state of $\mat{H}_M$, the resulting final state will converge to the ground state of $\mat{H}_P$~\cite{kremenetski2021quantum}. 

The linear schedule can be optimized by setting $\Delta$ to be a free parameter~\cite{shaydulin2021classical}. We denote this setting as the optimized linear schedule (OLS). Specifically, in QAOA with OLS, we fix the depth to a large value (e.g., $p=100$) and optimize $\Delta$ to minimize the energy:
\begin{equation}\label{eq:phase_diagram_optmize}
    \min_{\Delta} \bra{\vect{\psi}(\vect{\gamma}(\Delta),\vect{\beta}(\Delta))} \mat{H}_P \ket{\vect{\psi}(\vect{\gamma}(\Delta),\vect{\beta}(\Delta))}.
\end{equation}
Compared with~\eqref{eq:finite_optimize} with $2p$ variables, QAOA with OLS has only one free parameter $\Delta$ to optimize regardless of $p$, and hence is much easier to search for the optimum. Given the effectiveness of OLS in large-depth QAOA where regular QAOA parameter optimization becomes intractable~\cite{kremenetski2021quantum,kremenetski2023quantum}, we include QAOA with OLS in this study.

Adiabatic quantum algorithm (AQA).
AQA~\cite{Farhi2001,Albash2018} prepares the ground state of some target Hamiltonian by performing adiabatic evolution. Specifically, it proceeds from an initial Hamiltonian whose ground state is easy to prepare to a final Hamiltonian whose ground state encodes the solution to the computational problem. 

For a system evolving under a time-dependent Hamiltonian $\mat{H}(t)$, its time-evolution is governed by the Schr\"{o}dinger equation
\begin{equation}
    i\frac{\partial \ket{\vect{\psi}(t)}}{\partial t} = \mat{H}(t) \ket{\vect{\psi}(t)}.
\end{equation}
The quantum adiabatic theorem guarantees that if the initial state $\ket{\vect{\psi}(0)}$ is the ground state of $\mat{H}(0)$ and $\mat{H}(t)$ varies sufficiently slowly with $t$, the quantum state $\ket{\vect{\psi}(t)}$ will remain in the ground state of the instantaneous Hamiltonian $\mat{H}(t)$ for all $t$.

Connection between QAOA and AQA.
QAOA has important connections to AQA and its non-adiabatic variant. 
Ref.~\cite{kocia2022digital} shows that with a sufficiently large depth, QAOA with optimal angles can become a digitization of quantum annealing. 
Ref.~\cite{Brady2021} applies optimal control theory to solving the protocol for controlling the Hamiltonian evolution in both quantum annealing and QAOA. 
The optimal QAOA parameter schedule matches the optimal control protocol for AQA~\cite{brady2021behavior}.
Ref.~\cite{an2022quantum} shows that both AQA with tuned scaling and QAOA with an optimal control protocol can solve a quantum linear system problem.
An analog version of the QAOA by parameterizing and optimizing the schedule function is proposed in Ref.~\cite{barraza2022analog}.
Ref.~\cite{headley2022approximating} shows the possibility of running QAOA in a customized device with the digital analog paradigm.
Ref.~\cite{garcia2023lower} derives the lower bound of annealing time beyond the adiabatic regime.

While adiabatic evolution is a promising approach for quantum optimization, it suffers from potential non-adiabatic transitions between eigenstates of the system at time points where the Hamiltonian has small energy gaps and it is often infeasible for near-term devices due to noise and limited coherence times. Counterdiabatic driving is a method that compensates for the non-adiabatic effects by adding an additional term to the evolved Hamiltonian. 
The counterdiabatic evolution has been shown to improve adiabatic quantum optimization in~\cite{hegade2022digitized}.
In addition, the counterdiabatic term and counterdiabatic-inspired ansatz have also been found to benefit QAOA performance~\cite{hegade2022portfolio,chandarana2022digitized,chandarana2022digitized2,chai2022shortcuts}. 

Motivated by the connection between QAOA and AQA, the initial state $\ket{\vect{\psi}_0}$ is typically set to be the ground state of the mixing Hamiltonian $\mat{H}_M$~\cite{farhi2014quantum}. However, in many cases, either the ground state of $\mat{H}_M$ or $\mat{H}_M$ itself is difficult to implement exactly. The behavior in the adiabatic regime suggests that if $\mat{H}_M$ or the initial state is not implemented exactly (meaning that the initial state is not aligned with the ground state of $\mat{H}_M$), the QAOA performance may be affected. However, the impact of such alignment  on the performance of low-depth QAOA has received little attention to date.
In this work, we systematically study this alignment effect and demonstrate that it can significantly benefit QAOA performance far from the adiabatic limit, even in the low-depth regime. 

In this section, we describe the results from numerical simulations applying QAOA to ten portfolio optimization problem instances with the number of assets $N=6$. Unless otherwise specified, we set $p = 1, 2, 3$ for unrestricted QAOA given by Eq.~\eqref{eq:finite_optimize} and set $p = 100$ for QAOA with OLS~\eqref{eq:phase_diagram_optmize}. All the circuits were simulated using the \texttt{qiskit\_aer\_statevector} simulator. To optimize both QAOA parameter schedules, we use the BFGS optimizer built in the \texttt{Scipy}~\cite{virtanen2020scipy} package, running with multiple initial guesses (50-250 depending on the problem dimension). 

Given a solution $\mat{x}$ (portfolio selection) to the problem, we use \emph{approximation ratio} ($\mathrm{AR}$) to quantify the quality of the solution, defined as
\begin{equation}
  \mathrm{AR}(\mat{x}) = 
  \begin{cases}
  \frac{f(\mat{x})-f_{\rm max}}{f_{\rm min}-f_{\rm max}}, & \sum_i x_i = K, \\
  0, & \sum_i x_i \neq K,
  \end{cases}
\end{equation}
where $f_{\rm min}$ and $f_{\rm max}$ are the maximum and minimum value of $f(\mat{x})$ among all feasible portfolios, i.e.,
\begin{equation}
\begin{aligned}
    f_{\rm min} = \min_{\sum_i {x_i} = K} f(\mat{x}), \\
    f_{\rm max} = \max_{\sum_i {x_i} = K} f(\mat{x}).
\end{aligned}
\end{equation}

\subsection{Alignment Effect with Exact Mixers}~\label{sec:result_exact_mixer}
To investigate the alignment effect between the initial state and the ground state of $\mat{H}_M$, we conduct numerical simulations comparing circuits with different pairs of initial states and exact mixers. 
Our simulations studied various $\XY$ mixers, including the exact ring-$\XY$ mixer, complete-$\XY$ mixer, and arbitrary mixers that will be explained later. The exact mixers are implemented by a unitary operator constructed from directly exponentiating the corresponding mixing Hamiltonian. Correspondingly, we prepare the initial state by assigning it as the ground state of a mixing Hamiltonian. 

Exact ring-$\XY$ and complete-$\XY$ mixers. 
We first look at the exact ring and complete mixers for comparison. We separate the results by the combination of initial state and mixer type, and we label such combinations by ``${\text{S}}$-$\text{H}$'' pairs. For example, we use ${\text{S}_{\text{complete}}}$-$\text{H}_{\text{ring}}$ to denote that the initial state is the ground state of a complete-$\XY$ mixing Hamiltonian whereas the mixing Hamiltonian is a ring-$\XY$ model. %
\begin{figure}[t]
    \centering
    \includegraphics[width = 3.3in]{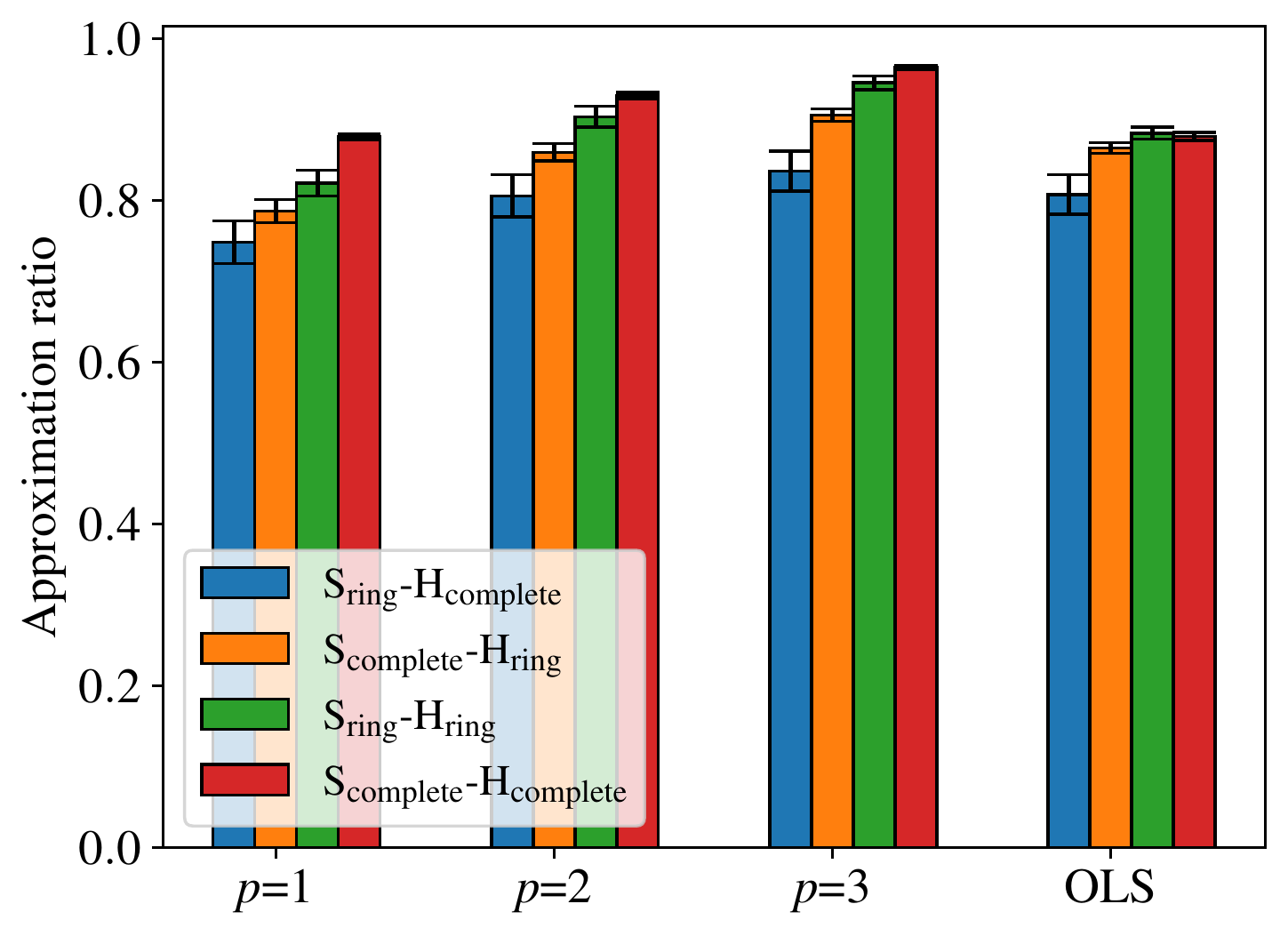}
    \caption{Comparisons of the exact ring- and complete-$\XY$ mixers in QAOA with unrestricted optimization at $p = 1, 2, 3$ and with the OLS method at $p = 100$. We reported the mean approximation ratios over $10$ instances with $N=6$ and $K=3$. The error bars represent standard errors of the mean. The alignment enhances performance in both low and high-depth QAOA.}
    \label{fig:completering_exact}
\end{figure}
\begin{figure}[t]
    \centering
    \includegraphics[width = 3.3in]{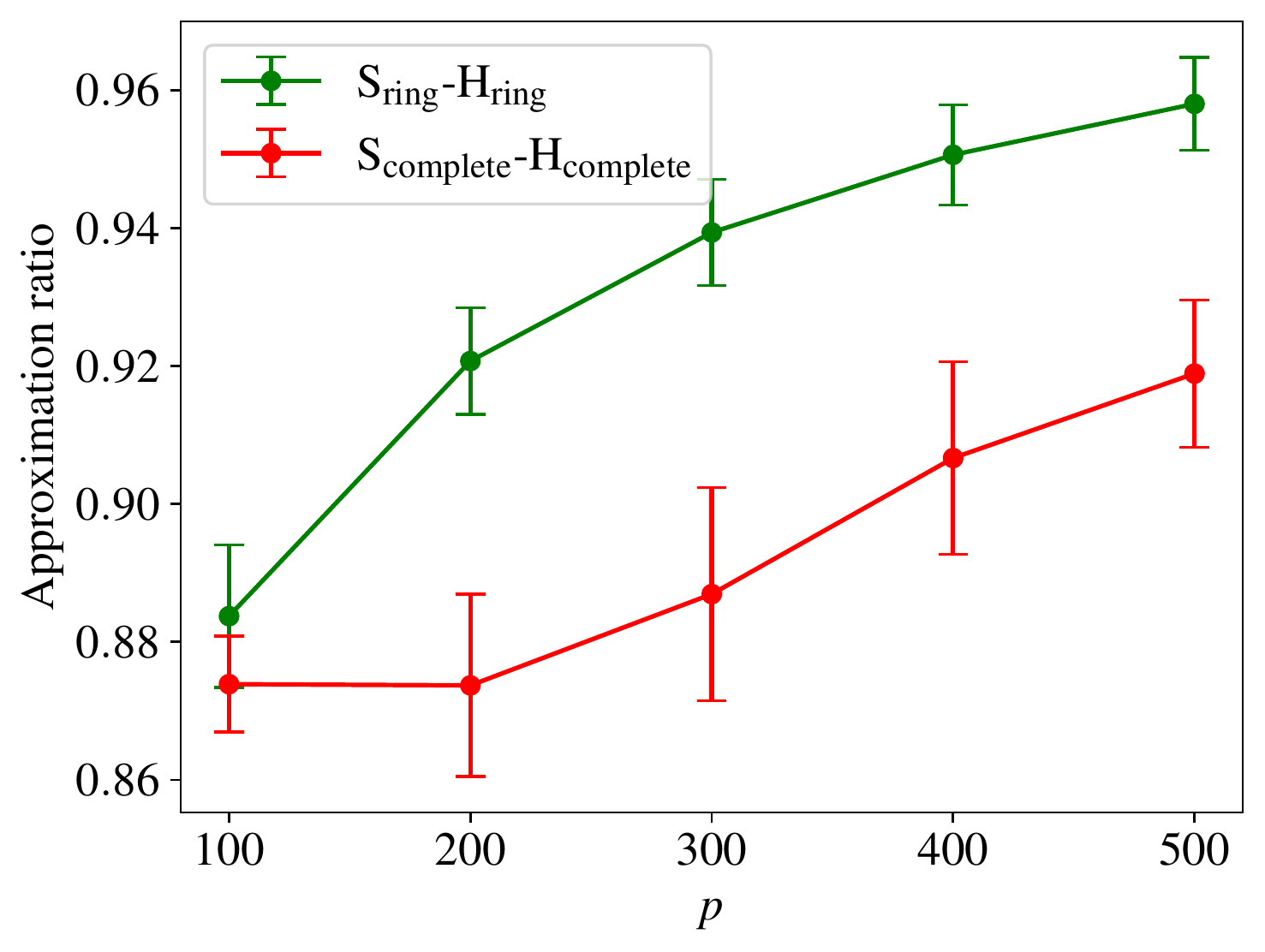}
    \caption{An example demonstrating the convergence of the OLS method for QAOA with exact mixers with the instances from Fig.~\ref{fig:completering_exact}. For both mixers, the initial states are aligned. As the QAOA depth increases, the final state of the OLS method~\eqref{eq:phase_diagram_optmize} will gradually converge to the ground state of the problem Hamiltonian $\mat{H}_P$. The complete-$\XY$ mixer needs a larger depth to converge with the OLS schedule. The error bars represent standard errors of the mean approximation ratios.
    }
    \label{fig:phase_diagram_converge}
\end{figure}
\begin{figure}[t]
    \centering
    \includegraphics[width = 3.3in]{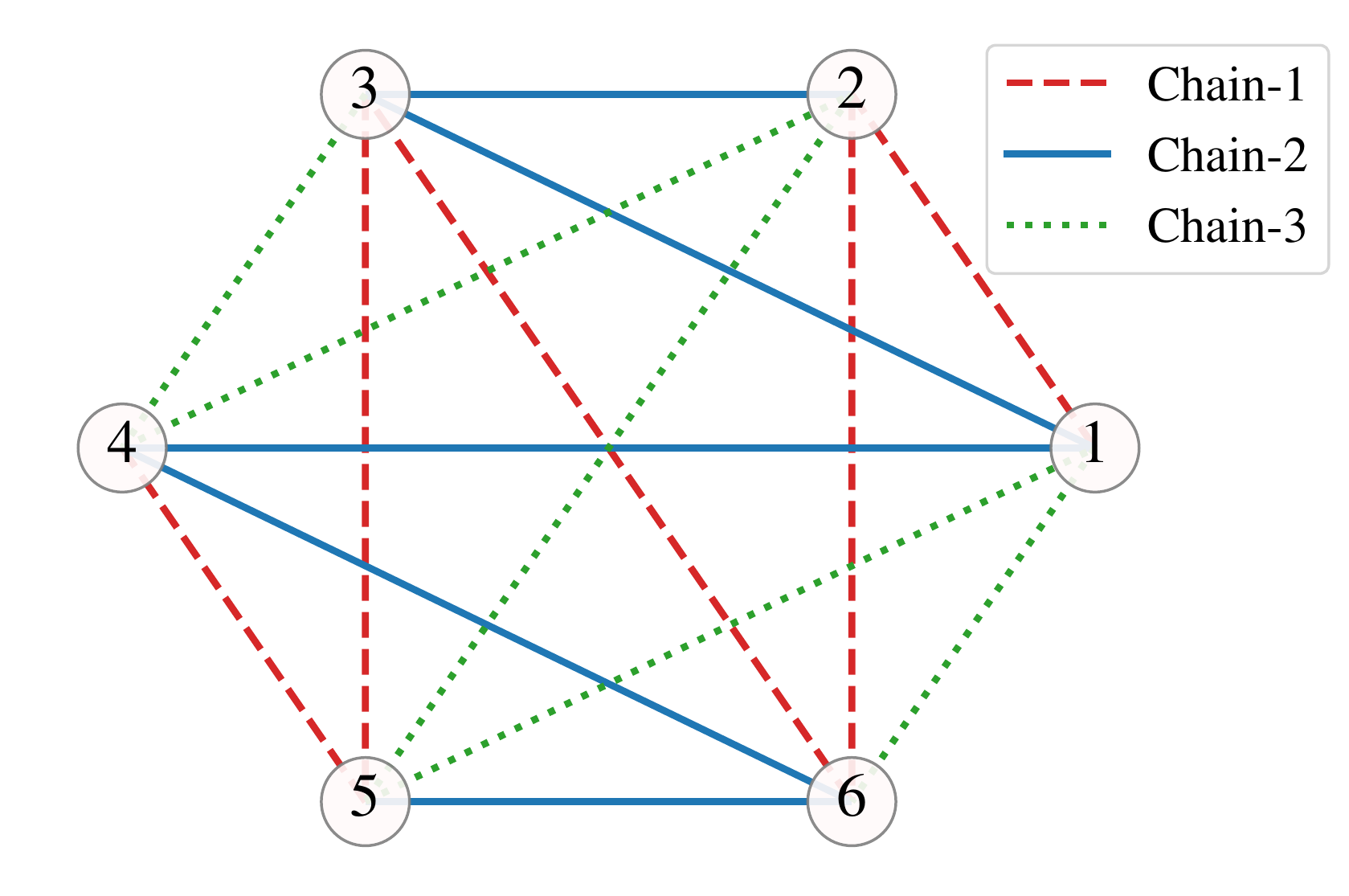}
    \caption{An example of the six-qubit complete-$\XY$ model: The complete graph is constructed by three separate chains, denoted by different colors and line styles. 
    }
    \label{fig:intermediadte_mixer}
\end{figure}
\begin{figure*}[t]
    \centering
    \includegraphics[width = 6.6in]{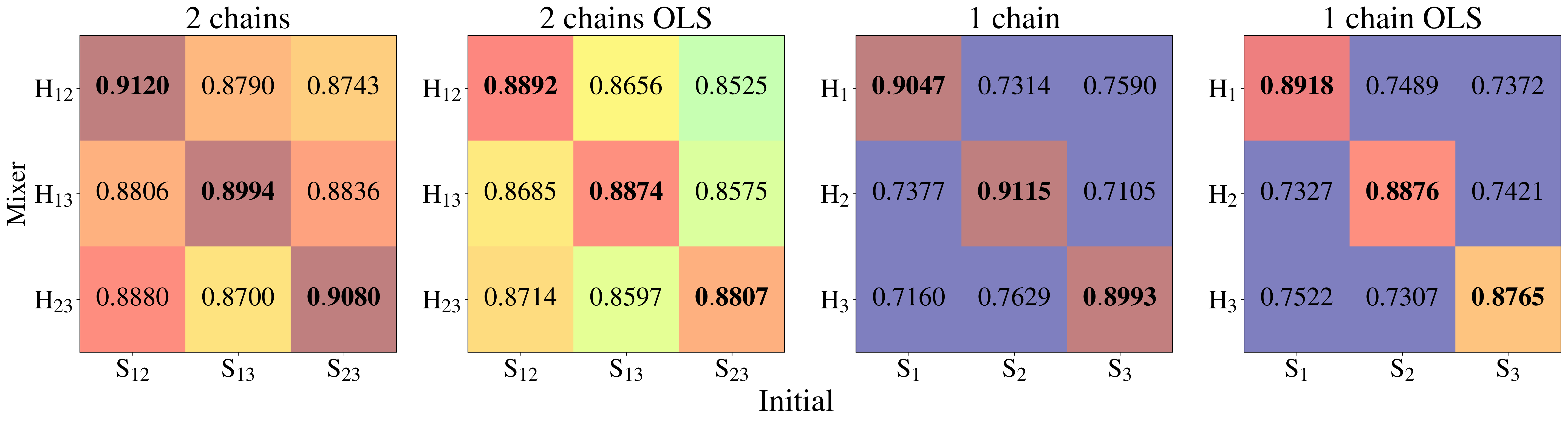}
    \caption{Comparisons of exact $\XY$-mixers in QAOA with unrestricted optimization at $p=2$ and with OLS at $p = 100$. The heatmaps display the average $\mathrm{AR}$ over the $10$ instances considered with $N=6$ and $K=3$. The mixers are constructed using one or two chains, as shown in Fig.~\ref{fig:intermediadte_mixer}. The alignment improves performance in both low and high-depth QAOA, as the diagonal pairs outperform others in the corresponding row and column.
    }
    \label{fig:partialchains_exact}
\end{figure*}

As shown in Fig.~\ref{fig:completering_exact}, for all $p$ studied, the ${\text{S}_{\text{complete}}}$-$\text{H}_{\text{complete}}$ pair gives significantly better $\mathrm{AR}$ than the ${\text{S}_{\text{complete}}}$-$\text{H}_{\text{ring}}$ pair, which aligns with the results reported in a previous study~\cite{cook2020quantum}. Similarly, the ${\text{S}_{\text{ring}}}$-$\text{H}_{\text{ring}}$ pair performs better than the ${\text{S}_{\text{ring}}}$-$\text{H}_{\text{complete}}$ pair. This indicates that alignment between the initial state and the ground state of $\mat{H}_M$ improves the QAOA performance for these cases, enabling the algorithm to converge more effectively to high-quality solutions. 

We note that it is not meaningful to directly compare the solution $\mathrm{AR}$ given by the unrestricted QAOA~\eqref{eq:finite_optimize} with that from QAOA with OLS~\eqref{eq:phase_diagram_optmize} since they have different depths and parameter schedules. The OLS method will gradually converge to the global optimum with a high enough depth (as shown in Fig.~\ref{fig:phase_diagram_converge}). The performance of the linear schedule could be less regular at a relative small $p$ (like $p=100$ and $200$ for ${\text{S}_{\text{complete}}}$-$\text{H}_{\text{complete}}$), which is referred as the ridge region in~\cite{kremenetski2023quantum}). For more detailed discussions on the linear parameter schedule, we refer the readers to Ref.~\cite{kremenetski2023quantum}. We only show the $p = 100$ results in Fig.~\ref{fig:completering_exact} and in the following sections as a sanity check to demonstrate that the alignment effect holds in QAOA with both a low and high depth. 
We also note that the performance improvement does not result from the warm-start effect, as the ${\text{S}_{\text{complete}}}$-$\text{H}_{\text{complete}}$ pair consistently outperforms the ${\text{S}_{\text{ring}}}$-$\text{H}_{\text{complete}}$ pair, and the ${\text{S}_{\text{ring}}}$-$\text{H}_{\text{ring}}$ pair also consistently performs better than the ${\text{S}_{\text{complete}}}$-$\text{H}_{\text{ring}}$ pair. 
This means that given different mixers, there is no such a fixed best initial state.

\subsection{Exact mixers with arbitrary connectivity}  
Next, we investigate the impact of alignment on some $\XY$ mixers that have arbitrary connectivity beyond ring and complete. $\XY$ models can be viewed as graphs with edges $(i,j)$ representing the indices $(i,j)$ of the interacting qubits. To satisfy the Hamming wight constraint of a solution, we have significant freedom to select edges and construct different variants of $\XY$ model. Here, we introduce an option for constructing mixers by selecting chains. 

For a complete-$\XY$ Hamiltonian with $N$ qubits (suppose $N$ is even), we can decompose the interaction terms as a summation of ${N}/{2}$ chains: 
\begin{equation}
    \mat{H}_{S_{\rm complete}}^{\XY} = \sum_{v=1}^{N/2} \mat{H}_{C_v}^{\XY} = \sum_{v=1}^{N/2}\sum_{(i,j)\in {C_v}} \mat{X}_{i}\mat{X}_{j} + \mat{Y}_{i}\mat{Y}_{j},
\end{equation}
where ${C_v}$ is the set of qubit indices in a chain. Inspired by the above decomposition, we can construct different $\XY$ mixing Hamiltonians by selecting a subset of chains from the complete graph. Notably, there are many possible ways to decompose a complete graph into chains, each of which may have different implications for QAOA performance. In our simulations, we arbitrarily select a decomposition of a $6$-node complete graph into $3$ chains as shown in Fig.~\ref{fig:intermediadte_mixer}. This approach allows us to compare the performance of QAOA with and without initial-mixer alignment across a range of $\XY$ mixers. We expect our conclusions to hold for other $\XY$ mixers as well.

Fig.~\ref{fig:partialchains_exact} illustrates the comparison of results from $\XY$ mixers constructed using the decomposition in Fig.~\ref{fig:intermediadte_mixer}, with the $x$-axis and $y$-axis indicating the initial state and mixing Hamiltonian labels, respectively. For example, the label 
${\text{S}_{\text{12}}}$ on the $x$-axis indicates that the initial state is the ground state of the mixing Hamiltonian built with chain-$1$ and chain-$2$, while the label ${\text{H}_{\text{12}}}$ on the $y$-axis indicates that the mixing Hamiltonian is built with chain-1 and chain-2. Our results show that the $\mathrm{AR}$s of diagonal pairs (i.e., with initial-mixer alignment) are significantly better than the non-diagonal pairs (without initial-mixer alignment), complementing the results observed for the exact ring-$\XY$ and complete-$\XY$ mixers. It suggests that the alignment effect applies to a wide range of $\XY$ mixers. Specifically, we found that the alignment effect is more pronounced for simpler mixers with less connectivity, such as those constructed using a single chain.

\begin{figure*}[t]
    \centering
    \includegraphics[width = 6.6in]{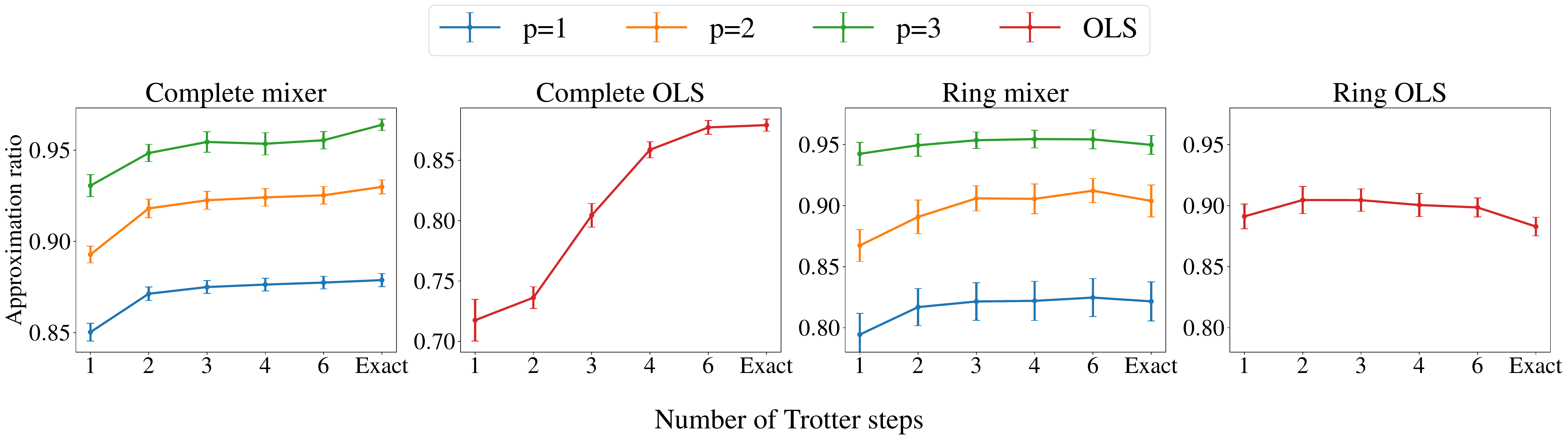}
    \caption{Comparisons of Trotterized ring- and complete-$\XY$ mixers in QAOA with unrestricted optimization at $p = 1, 2, 3$ and with the OLS method at $p = 100$. We report the mean approximation ratio over $10$ instances with $N=6$ and $K=3$ with error bars denoting the standard errors of the mean estimation.
    A larger Trotter number consistently results in better performance for the complete mixer. However, for QAOA with an approximated ring-$\XY$ mixer, we observe a more robust QAOA performance.
    }
    \label{fig:completering_T}
\end{figure*}
\begin{figure}[t]
    \centering
    \includegraphics[width = 3in]{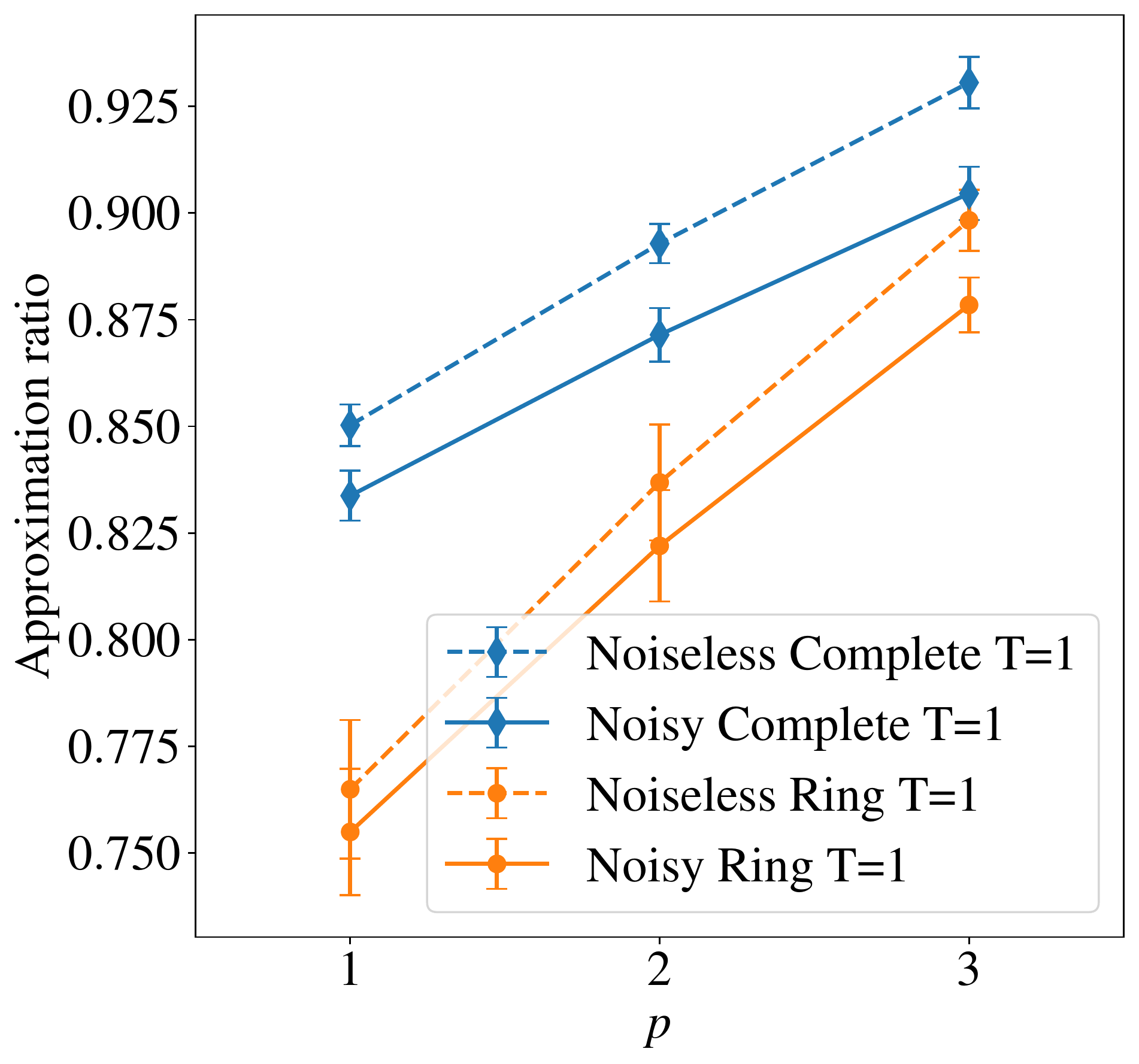}
    \caption{Given an initial Dicke state, the comparisons of Trotter-step-1 approximated ring- and complete-$\XY$ mixers in QAOA at $p = 1, 2, 3$. We report the mean approximation ratio over $10$ instances with $N=6$ and $K=3$ with error bars denoting the standard errors of the mean estimation. Under the noisy simulation, the complete-$\XY$ mixer still outperforms the ring-$\XY$ mixer, which demonstrates the alignment effect.
    }
    \label{fig:emulator_N6}
\end{figure}
\begin{figure*}[t]
    \centering
    \includegraphics[width = 6.6in]{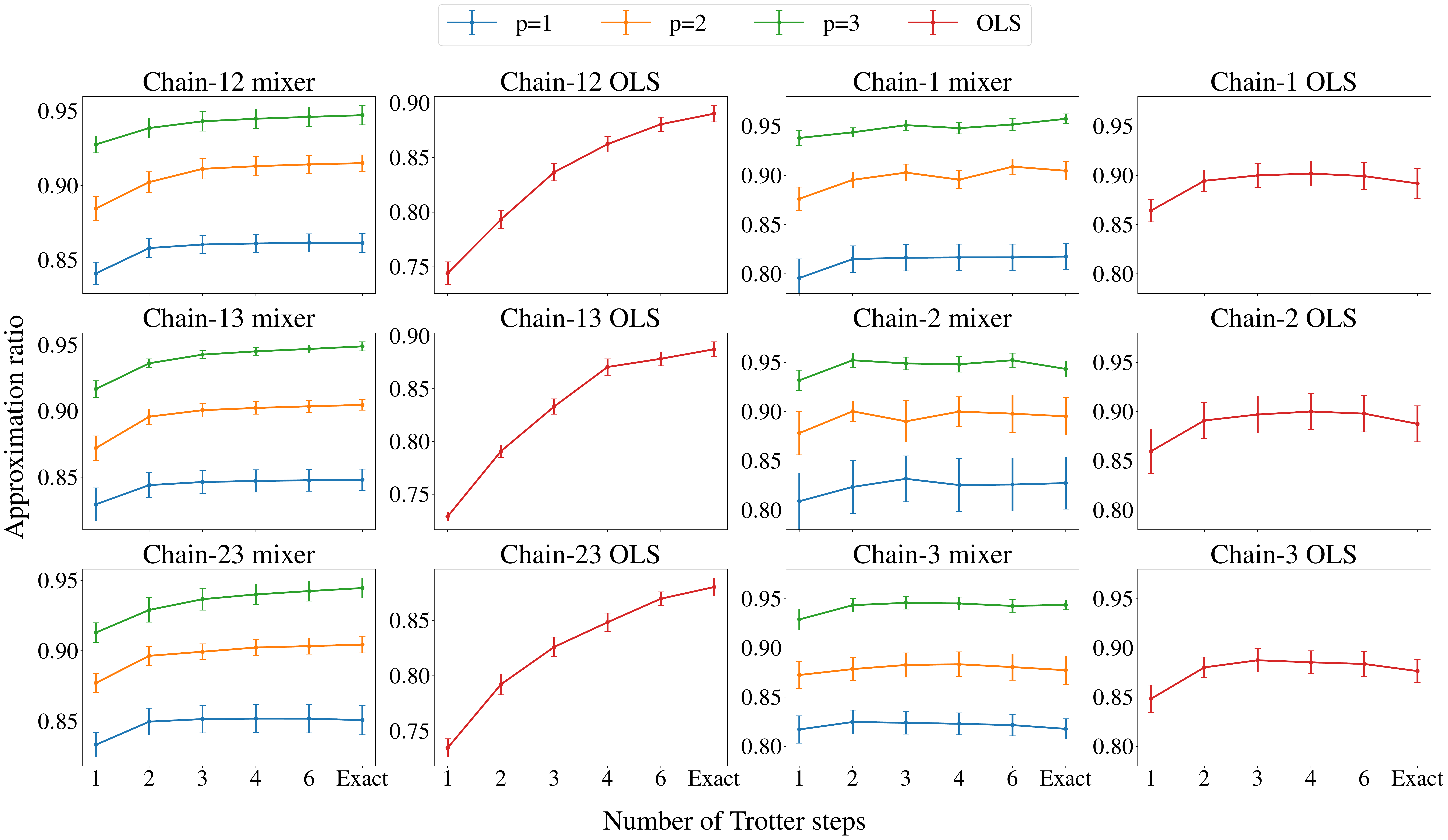}
    \caption{Comparisons of Trotterized $\XY$ mixers in QAOA with unrestricted optimization at $p = 1, 2, 3$ and with OLS method at $p = 100$. We report the mean approximation ratio over $10$ instances with $N=6$ and $K=3$ with error bars denoting the standard errors of the mean estimation. A larger Trotter step consistently leads to better performance for mixing Hamiltonians built with two chains. In contrast, for the approximated mixing Hamiltonians built with one chain, the performance improves when moving from Trotter step 1 to 2 and then stabilizes.}
    \label{fig:partialchains_T}
\end{figure*}
\subsection{Alignment Effect with Trotterized Mixers}~\label{sec:result_trotter_mixer}
Next, we explore the alignment effect in the practical circuits. To achieve it, we must decompose the mixing operator $e^{-i\beta \mat{H}_M}$ into a series of $1$-qubit and $2$-qubit gates. One of the widely used approaches is Trotterization. In the following, we will first describe the Trotterization procedure for various $\XY$ mixers, and discuss how the resulting Trotter error can impact the quality of the solution.

Mixer Trotterization. For ease of notation, we define ${\mat{\XY}}_{i} = \mat{X}_{i}\mat{X}_{j} + \mat{Y}_{i}\mat{Y}_{j}$, where $j = (i+1) \mod N$. 
For a one-dimensional (i.e., chain or ring) mixer, we use the popular parity partition strategy to Trotterize it: 
\begin{equation}\label{eq:trotter_singlechain}
    {e}^{-i\beta \sum_{i} {\mat{\XY}}_{i}} \approx {\left[ \prod_{j~\text{is odd}} e^{-i\frac{\beta}{T} {\mat{\XY}}_{j}} \prod_{j~\text{is even}} e^{-i\frac{\beta}{T}{\mat{\XY}}_{j}} \right]}^T,
\end{equation}
where $T$ denotes the number of Trotter steps, called the \emph{Trotter number}. 

When constructing a mixer using multiple $\XY$ chains, we apply a two-level approximation strategy. Firstly, we Trotterize the chains in sequential order with a Trotter number ${T_1}$. Secondly, we apply the parity partition strategy within each chain with a Trotter number ${T_2}$. Given a mixing Hamiltonian constructed by $k$ chains, $\sum_{v=1}^k \mat{H}_{\rm{C}_v}^{\XY}$, we approximate its unitary as follows:
\begin{equation}\label{eq:trotter_multiplechain}
\begin{aligned}
    &{e}^{-i\beta \sum_{v=1}^k \mat{H}_{\rm{C}_v}^{\XY}} \approx {\left[\prod_{v=1}^k {e}^{-i\frac{\beta}{T_1} \mat{H}_{\rm{C}_v}^{\XY} }\right]}^{T_1} \quad \rm{with}\\
    &{e}^{-i\frac{\beta}{T_1} \mat{H}_{\rm{C}_v}^{\XY} } \approx {\left[ \prod_{j=\rm{odd}} e^{-i\frac{\beta}{T_1 T_2} {\mat{\XY}}_{j}} \prod_{j=\rm{even}} e^{-i\frac{\beta}{T_1 T_2} {{\XY}}_{j}} \right]}^{T_2}.
\end{aligned}
\end{equation}

The choices of $T_1$ and $T_2$ control the Trotter error. Given a Hamiltonian $\mat{H}=\mat{H}_1+\mat{H}_2$ with evolution time $t$, the commutator-type error bound for its first-order Trotter approximation is as follows~\cite{lin2022lecture}: 
\begin{equation}
    \| {e}^{i t\mat{H}} - {e}^{i t \mat{H}_1}{e}^{i t \mat{H}_2} \|_{2} \leq \frac{t^2}{2}\| \left[ \mat{H}_1, \mat{H}_2\right] \|_{2}.
\end{equation}
Intuitively, the spectral norm of the commutator between two chain Hamiltonians $\mat{H}_{\rm{C}_v}^{\XY}$ will be significantly larger than the one between the two parity-partitioned parts of one chain Hamiltonian. 
Therefore, in our implementation, we fix $T_2=1$ and adjust $T_1$ to control the approximation accuracy.
Generally, a $T$-step Trotterization will enlarge the mixer circuit for $T$ times. However, using a large Trotter number can be computationally costly, even in simulation. Therefore, we implement the Trotterized mixer operators with steps up to six, which we find is sufficient for our purposes. 

In the following simulations analyzing the impact of alignment on QAOA performance, we will fix the initial state as the ground state of the exact mixing Hamiltonian and approximate the mixer operator via different Trotter numbers. While some $\XY$ mixers can be implemented exactly, such as the ring-$\XY$ mixer can be realized by diagonalization~\cite{verstraete2009quantum} or other algebraic compression techniques~\cite{gulania2022quybe}, and a $2^m$-sized complete-$\XY$ mixer (with $m$ being a positive integer) can be realized efficiently for a Hamming-weight $K=1$ problem~\cite{wang2020xy}, we chose to use the Trotterization to implement all the mixers in our simulations. The reason for this choice is that the Trotterization is more flexible and allows us to analyze the QAOA performance under mixers with various approximation accuracies more easily. It is worth noting that there also exist other Trotter strategies~\cite{tranter2019ordering,lin2022lecture,childs2021theory}, which are out of the scope of this paper. 

Trotterized ring-$\XY$ and complete-$\XY$ mixers.
First, we focus on the Trotterized ring-$\XY$ and complete-$\XY$ mixers.
Fig.~\ref{fig:completering_T} shows the QAOA performance under different approximated mixers with low and high QAOA depths. 
In the case of the complete-$\XY$ mixer, an increase in the Trotter number consistently enhances QAOA performance, converging to the ${\text{S}_{\text{complete}}}$-${\text{H}_\text{complete}}$ results depicted in Fig.~\ref{fig:completering_exact}. For the ring-$\XY$ mixer, QAOA performance exhibits greater robustness in terms of the Trotter number. 
However, an initially more accurate mixer continues to contribute to performance improvement, such as the performance observed when increasing Trotter number from $1$ to $2$. 
The results of low-depth QAOA parameter optimization and high-depth linear schedule simulations show a consistent relationship, specifically, their $\mathrm{AR}$ results follow the same trend with respect to Trotter number.

We hypothesize that the distinct behavior observed between Trotterized ring-$\XY$ and complete-$\XY$ mixers arises from the intricacy of their respective mixing structures. To substantiate this hypothesis, we conduct subsequent numerical experiments employing Trotterized variants of $\XY$ mixers.

We also demonstrate the alignment effect in noisy simulation via the Quantinuum's H2-1 device emulator. Considering the practical circuit implementation, we prepare the circuits with the Dicke state (a uniform superposition over bitstrings with a fixed Hamming weight) and Trotter-step-1 approximated ring-$\XY$ and complete-$\XY$ mixers. We report both the noisy and noiseless simulation results in Fig.~\ref{fig:emulator_N6}. We observe that for this problem with the total number of $2$-qubit gates less than $200$, given an initial Dicke state, the alignment effect still holds where the complete-$\XY$ mixer achieves better performance than the ring-$\XY$ mixer in the presence of realistic noise.

Trotterized arbitrary mixers.
Next, we analyze the alignment effect for the Trotterized $\XY$ mixers. To investigate the impact of mixer structure, we did the same simulations for six  $\XY$ mixers, including three whose Hamiltonians are built with two chains and three whose Hamiltonians are built with one chain. As illustrated in Fig.~\eqref{fig:partialchains_T}, we observe consistent trends between the $2$-chain mixers and the complete-$\XY$ mixer, as well as between the one-chain mixers and the ring-$\XY$ mixer. Based on these observations, we argue that when the initial state aligns with the ground state of the exact mixing Hamiltonian, a more precise implementation of the mixing Hamiltonian with complex connectivity leads to improved performance. In contrast, for a less connected mixing Hamiltonian, a Trotterized implementation with a few steps attains optimal performance, which then stabilizes.

In summary, our results demonstrate that the alignment effect positively impacts QAOA performance across both low- and high-depth regimes. This observation was validated through the application of exact and Trotterized mixers on various $\XY$ mixers.

\subsection{Experiments on a Trapped-ion Quantum Processor}
\begin{figure}[t]
    \centering
    \includegraphics[width = 3.3in]{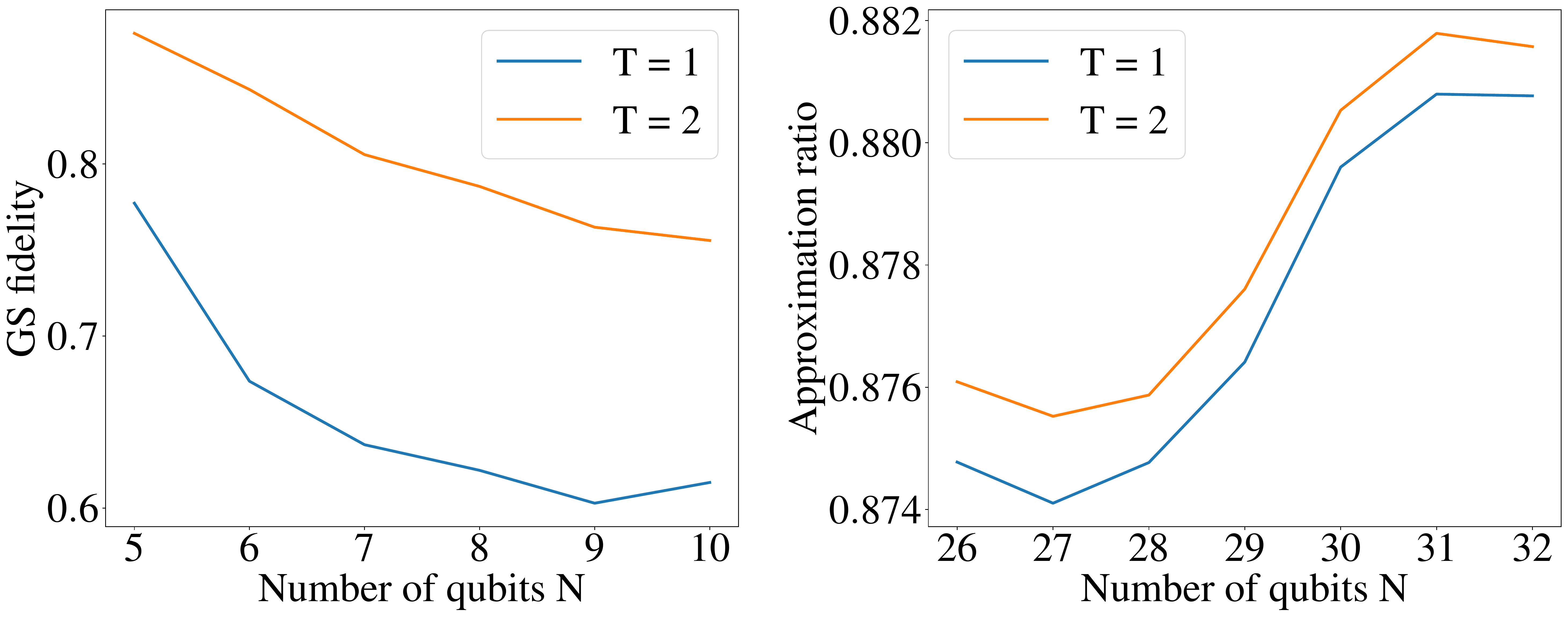}
    \caption{Left: an example of achieving a larger overlap between the Dicke state and the effective ground state of the Trotterized ring mixer at $\beta = 0.5$ by increasing the Trotter number from $1$ to $2$. The Dicke state is prepared with different $N$ values but a fixed $K=3$. Right: the quality of QAOA solution in noiseless simulation. The initial state is prepared as the Dicke state with a fixed $K=5$, and the ring-$\XY$ mixer is approximated with Trotter numbers $1$ and $2$. For various problem sizes, we consistently observe a performance improvement when transitioning the Trotter number from $1$ to $2$. 
    }
    \label{fig:dicke_fidelity}
\end{figure}

While the improvement in performance from alignment between the initial state and the ground state of $\mat{H}_M$ is robust in the noiseless simulation, its absolute value is relatively small. Intuitively, this suggests that noise will likely affect it when executed on near-term hardware. 
We now demonstrate this intuition by executing QAOA with Trotterized ring-$\XY$ mixer on Quantinuum H2-1 trapped-ion processor using $32$ qubits. 

For the ring-$\XY$ mixer, the ground state of the exact Hamiltonian can be difficult to prepare in a quantum circuit, especially on noisy hardware. Therefore, in our experiments, we use the Dicke state as a proxy for the ground state of the exact ring-$\XY$ Hamiltonian, and use it as the initial state of the QAOA circuit as well as the target state in evaluating the overlap with the effective ground state. 
The Dicke state is prepared using the divide-and-conquer approach of~\cite{aktar2022divide}. Fig.~\ref{fig:dicke_fidelity} shows that increasing the Trotter number from $1$ to $2$ improves the fidelity between the Dicke state and the effective ground state, subsequently improving QAOA performance in the noiseless simulation. However, we do not expect this to hold strictly as the Trotter number increases, where an exact ground state would need to be prepared.

\begin{figure}[t]
    \centering
    \includegraphics[width = 3.3in]{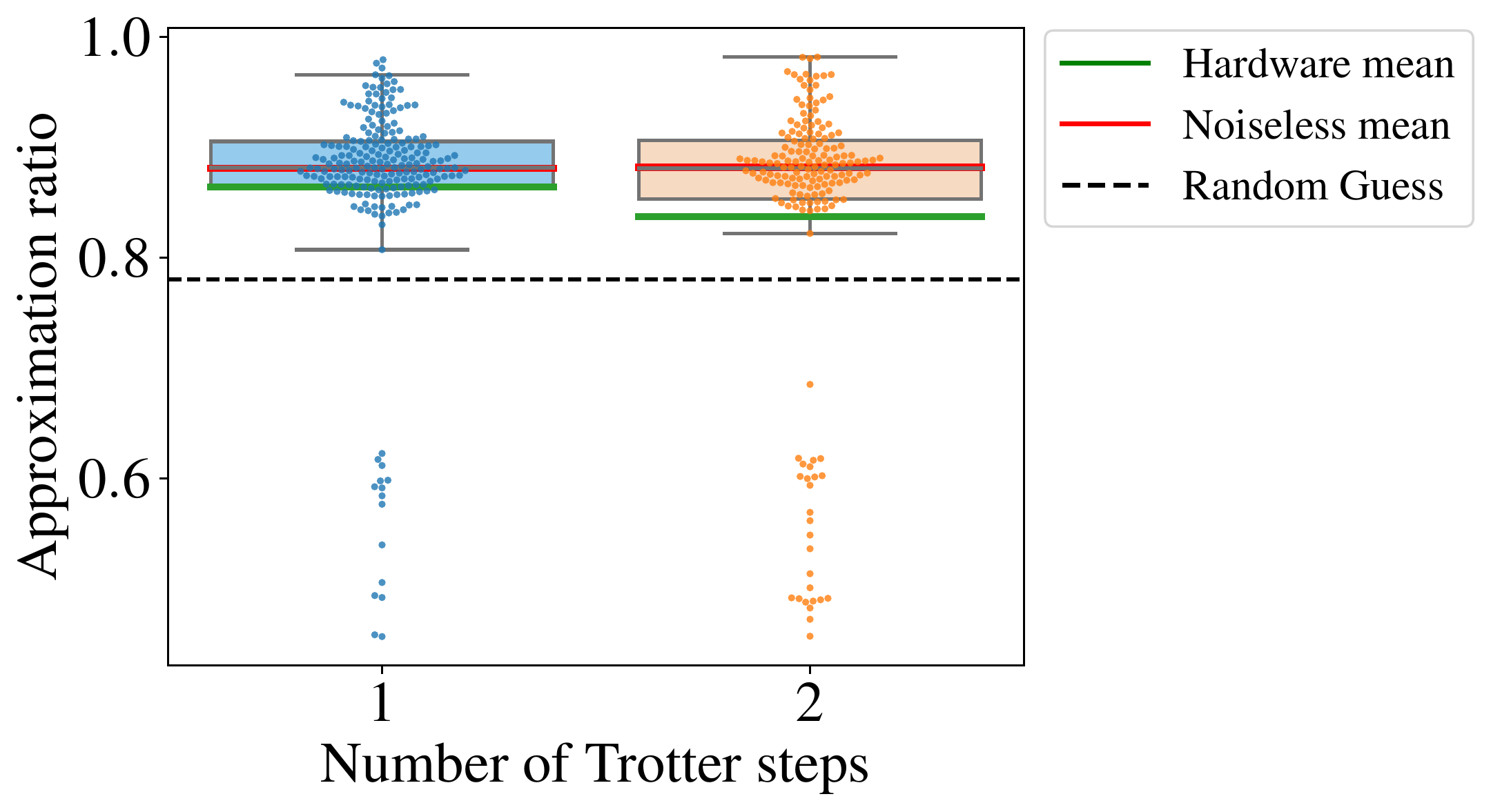}
    \caption{
    Experimental results for $N=32$ in trapped-ion quantum hardware: The ${\rm AR}$ from the random guess (uniform over all feasible solutions) is $0.7801$. For ${\rm ARs}$ from Trotter-number-$1$ ($T=1$) and Trotter-number-$2$ ($T=2$) QAOA, hardware results are $0.8638$ and $0.8424$, while noiseless simulator results are $0.8808$ and $0.8816$. The hardware results, though impacted by noise, are significantly better than the random guess. To evaluate ${\rm ARs}$, we post-selected feasible solutions in the hardware experiments, selecting $213$ and $172$ feasible samples out of $2500$ shots in $T=1$ and $T=2$ experiments. The post-selection ratio is significantly better than a random selection of $0.0047\%$. The p-value of the independent two-sample t-test for the two groups of feasible samples is $0.0344$, indicating a significant difference between the means of the two groups. For visualization purposes, we exclude one outlier with ${\rm AR}$ $<0.35$ in the swarm plot of both $T=1$ and $T=2$ experiments. 
    }
    \label{fig:N32_result}
\end{figure}

The QAOA circuit is compiled to H2-1 and optimized using pytket~\cite{Sivarajah_2020}, resulting in the total numbers of $2$-qubit gates of $1,159$ for $T=1$ and $1,223$ for $T=2$. We note that the Dicke state preparation needs $581$ $\cnotgate$ gates. As shown in Fig.~\ref{fig:N32_result}, we observe that in the hardware experiments, the performance of the $T=2$ circuit is worse than the $T=1$ one. It validates that the improvement of Trotter approximation can be impacted by the hardware noise. However, the hardware results are still significantly better than the random guess, shedding light on the power of advanced quantum devices. The hardware results can be further improved by performing error mitigation techniques such as symmetry verification by parity checks~\cite{Shaydulin2021EM,Kakkar2022,Gonzales2023EM}. 
\section{Discussion}\label{sec:discussion}
In this paper, we demonstrate that the alignment effect is impactful even at very small QAOA depth, suggesting a strong connection between QAOA and adiabatic quantum algorithms.
We show the evidence of the alignment effect by studying QAOA performance with various $\XY$ mixers in two ways: with the exact mixers and varying initial states, and with a fixed initial state and varying fidelity of the mixer implementation. 
We use portfolio optimization problems as the benchmark, but we expect the findings to apply broadly to other combinatorial optimization problems.
To the best of our knowledge, this is the first study of the impact of Trotter approximation error in mixer implementation on QAOA performance. 
For simple one-dimensional $\XY$ mixers, the QAOA performance is relatively robust to the Trotter error. Meanwhile, for the more complicated $\XY$ mixers, a larger Trotter number leads to better performance since the effective ground state is approaching the initial state (the ground state of the exact mixing Hamiltonian).

While we show that better alignment improves performance, for small system sizes accessible numerically 
the absolute value of the improvement is relatively small. 
For instance, in hardware experiments on the H2-1 device, we do not observe the anticipated improvement in solution quality when increasing Trotter number from $1$ to $2$. This highlights the centrality of minimizing the circuit depth when executing QAOA on NISQ devices.

Beyond demonstrating the alignment effect, designing constraint-preserving mixers is of independent interest~\cite{fuchs2022constraint,radha2021quantum}. In a recent paper~\cite{fuchs2022constraint}, the authors studied the mixer design from the perspective of a transition matrix. For a Trotterized $\XY$ mixer, some transitions between feasible states may be suppressed when the Trotter number is not large enough. However, our results show that QAOA performance is not explicitly related to the transition path. For instance, in the Trotterized complete mixers, all the possible transitions between feasible states have been filled with Trotter number one. Meanwhile, the Trotter number still greatly influences the QAOA performance, as shown in Fig.~\ref{fig:completering_T} and~\ref{fig:partialchains_T}. This underscores the importance of both the connectivity and probability of transitions in mixer design for achieving high performance in QAOA.

\section{Methods}\label{sec:methods}
In this section, we introduce and discuss some implementation details that enabled the simulations presented in the results section.

\subsection{Problem Instances Selection}\label{sec:instance_select}
We first generate a pool of portfolio optimization problem instances by randomly generating the mean return vector and covariance matrix using \texttt{RandomDataProvider} in \texttt{qiskit\_finance}~\cite{qiskitfinance}. To make the performance of different QAOA variants clearly distinguishable, we intentionally select the ``hard'' instances. These instances are chosen by roughly examining depth-1 QAOA performance with the initial state set to be the Dicke state and Trotter-step-1 approximated ring-$\XY$ and complete-$\XY$ mixers. Only instances that have relatively low $\mathrm{AR}$ (${\mathrm{AR} < 0.8}$ for ring-$\XY$ and ${\mathrm{AR} < 0.85}$ for complete-$\XY$) are included in the benchmark. We choose a total of $10$ instances as our benchmark: $5$ from QAOA with a Trotterized ring-$\XY$ mixer and $5$ from QAOA with a Trotterized complete-$\XY$ mixer. 

\begin{figure*}[t]
    \centering
    \includegraphics[width = 6.2in]{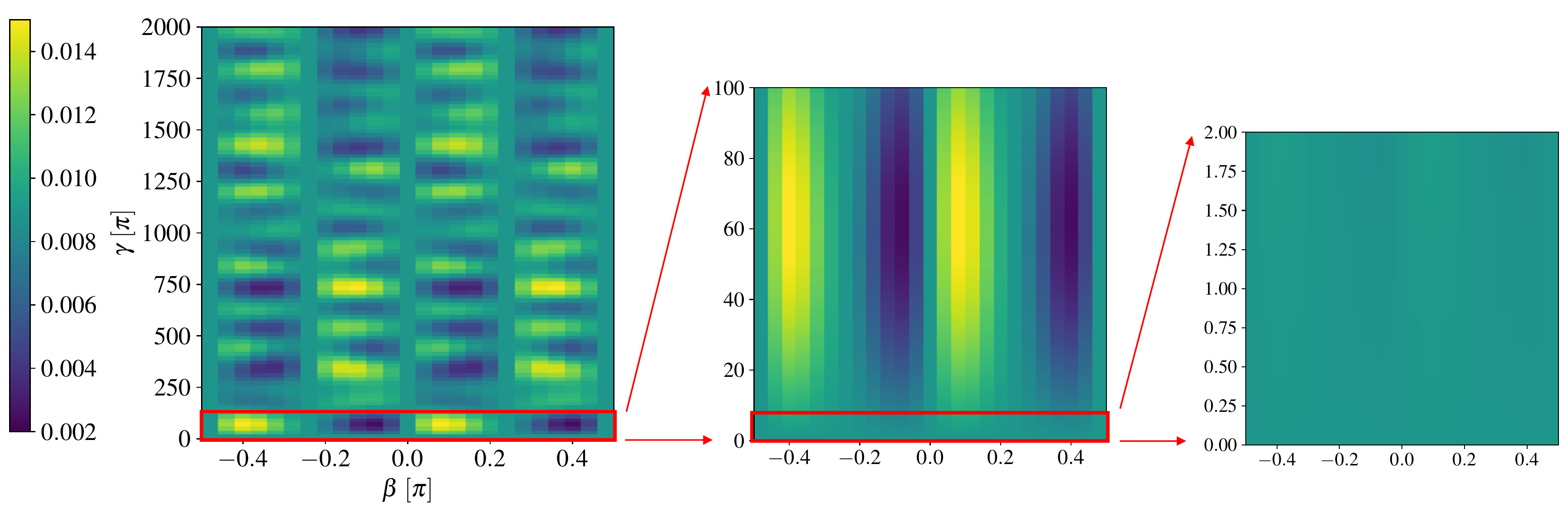}
    \caption{This example demonstrates how to select the rescaling factor for one $N=6$ instance. The figures from left to right display the depth-$1$ QAOA energy landscape with gamma search space $[0,2000\pi]$, $[0,100\pi]$, and $[0,2\pi]$. By applying a rescaling factor and fixing the search space as $[0,2\pi]$, they are equivalent to setting the rescaling factors as $1000$, $50$, and $1$, respectively. In this example, a rescaling factor of $50$ encompasses high-quality local minima in the landscape.
    }
    \label{fig:rescaling_factor_select}
\end{figure*}
\subsection{Improving the Trainability}\label{sec:improve_trainability}

One of the challenges in solving the portfolio optimization problem~\eqref{eq:portfolio_opt} with QAOA is that due to the non-integer weights (mean returns and covariances between assets) assigned to the problem Hamiltonian terms, the QAOA objective \eqref{eq:finite_optimize} is not periodic. %
A larger parameter search space will correspondingly require more initial points in a classical numerical optimizer to converge to a high-quality local optimum. Different orders of $\vect{\gamma}$ and $\vect{\beta}$, and consequently different orders of their gradients, can also introduce difficulties to the classical optimizer. 

To address it, we multiply the objective function~\eqref{eq:portfolio_opt} by a rescaling factor $\lambda$, which is a predefined instance-dependent constant. Such a rescaling factor does not influence the true solution to the problem, but it allows us to control the search range in the energy landscape and rescale the order of $\vect{\gamma}$ and its gradient. In QAOA, it is equivalent to scaling the $\vect{\gamma}$ to $\vect{\gamma}^\prime = \lambda \vect{\gamma}$. In a numerical optimizer, if we fix a bounded search range of $\vect{\gamma}$, such as $ \left[0,2\pi\right]^p$, scaling by $\lambda$ is equivalent to extending the search range to $\left[0,2\lambda \pi\right]^p$.
In general, we are not guaranteed to find a global optimum in this fixed interval; in fact, adversarial examples can be constructed with a global optimum far from origin~\cite{shaydulin2022parameter}. However, in practice, we observe that this technique always gives a high-quality local optimum.
In this paper, we use the following protocol to select the rescaling factor $\lambda$: 
\begin{itemize}
    \item Implement a QAOA with $p = 1$ for the specific problem instance.
    \item Plot the heatmap of the circuit performance by conducting a grid search over $\gamma$ and $\beta$ in a bounded range (e.g., $\left[0,2\pi\right]$ and $\left[-\frac{\pi}{2}, \frac{\pi}{2} \right]$). A coarse search grid may be used for efficiency.
    \item Select the rescaling factor $\lambda$ by controlling the range of $\gamma$ to cover the regions on the heatmap with high-quality local optima.  
\end{itemize}
An example of selecting the rescaling factor is depicted in Fig.~\ref{fig:rescaling_factor_select}. The selection of $\lambda$ is not sensitive to the circuit structures discussed in the results section. Similar protocols for rescaling the QAOA objective have been proposed in Refs.~\cite{boulebnane2022peptide,brandhofer2023benchmarking,shaydulin2022parameter}. Similarly to previous results, we observe that using one rescaling factor for all $p$ works well.
When the problem size is large, we can make use of some advanced QAOA simulators to obtain the expected energy, such as~\cite{lykov2022tensor,mandra2021hybridq,ibrahim2022constructing}. 

To study the alignment effect with Trotterized mixers, we try to explore the performance within the same landscape. To avoid the numerical optimizer driving the solution out of the targeted landscape, we fix the same search range of $\vect{\beta}$ for all Trotterized implementations and set a hard boundary constraint on the solved $\vect{\gamma}$. In the regular QAOA parameter optimization, for the circuits with the same setup but different Trotter numbers, a solution from one circuit could be a good initial guess for other circuits. 

\subsection{Initial State Preparation}~\label{sec:initial_state}
Circuit realization. In our alignment effect simulations, we need to prepare the initial state as the ground state of the corresponding mixing Hamiltonian. However, in general, the state preparation circuit for constructing the initial state can be costly to implement. 
In our implementation, we skip the gate-based circuit realization by assigning an exact state vector as the initial state in the simulator. One special case is the complete-$\XY$ mixer. The ground state of its Hamiltonian is the Dicke state, whose efficient circuit implementation is known~\cite{aktar2022divide,bartschi2022short} but could still be costly in the near term devices. For example, Ref.~\cite{aktar2022scalable} studies fidelity lower bounds of Dicke state preparation on Quantinuum H1 devices.

\begin{figure}[t]
    \centering
    \includegraphics[width = 3.3in]{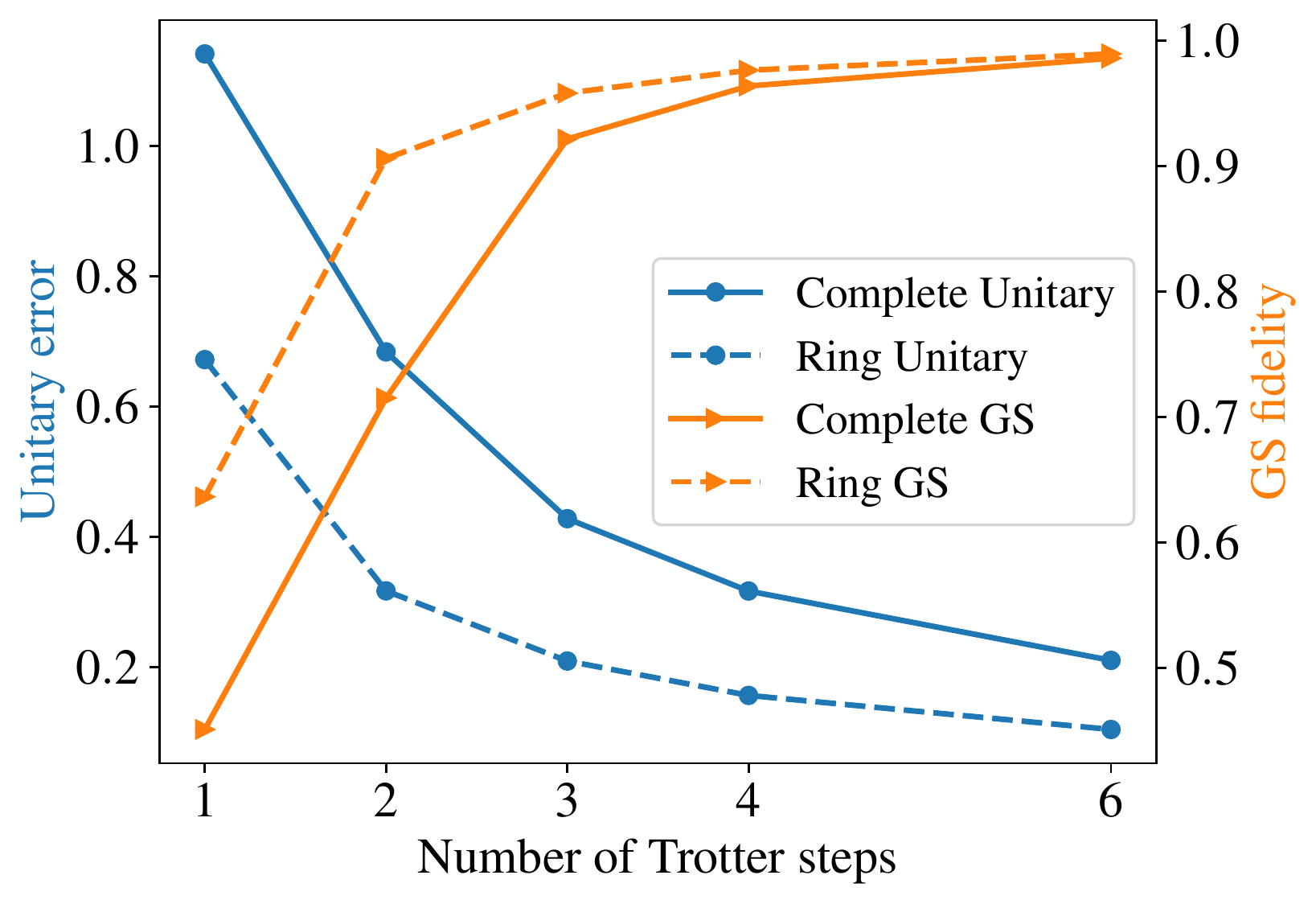}
    \caption{An example of Trotterizing the unitary $\mat{U} = e^{-i \beta \mat{H}_{S}^{\XY}}$ with $N$=6 using different numbers of step with $\beta=0.5$: The blue lines represent the relative error in approximating the unitary, while the orange lines depict the fidelity between the ground states of exact and effective Hamiltonians.}
    \label{fig:unitary_gsfidelity_T}
\end{figure}

\subsection{Determination of the ground state of the mixing Hamiltonian}\label{sec:effective_H}
In the case of the mixer implemented exactly, we can determine the ground state by numerically performing the eigenvalue decomposition on the mixing Hamiltonian. Since we are considering a Hamming weight constraint, we only consider the ground state in the feasible subspace.

However, when the mixer is implemented using Trotter approximation, statements about the spectral properties of the exact mixer may not be valid. Trotterization can make notions like the ``ground state of the mixing Hamiltonian'' ambiguous. Specifically, if a mixer $\mat{U}(\beta)$ is the product of non-commuting operators, its eigenvectors, and consequently the eigenvectors of any Hermitian operator $\mat{H}(\beta)$ such that $\mat{U}(\beta) = e^{-i\mat{H}(\beta)}$, become dependent on $\beta$. 
For this reason, we prepare the initial state as the ground state of the exact mixing Hamiltonian and try to approximate the mixer with a larger Trotter number $T$. 

In addition, even for a fixed $\beta$, the periodicity of the eigenvalues of $\mat{U}(\beta)$ (which are all unit complex numbers) allows each eigenvalue of $\mat{H}(\beta)$ to be shifted by a multiple of $2\pi$, while still corresponding to the same eigenvalue and eigenvector of $\mat{U}(\beta)$. This phenomenon is also discussed in~\cite{kremenetski2023quantum}. It renders the notion of the ``smallest eigenvalue and the associated eigenstate'' ill-defined.
To quantify the alignment level between the initial state and the mixer, similarly to Ref.~\cite{yi2021success}, we define the \emph{effective Hamiltonian} for a Trotterized unitary operator and its associated \emph{effective ground state}, based on the intuition from the adiabatic limit. Specifically, the effective Hamiltonian associated with evolution time $\beta$ is defined as $\mat{H}_\text{eff}({\beta}) = i\log(\mat{U}(\beta))$ and the corresponding effective ground state is the eigenstate of $\mat{H}_\text{eff}({\beta})$ that exhibits maximal overlap with the ground state of the exact Hamiltonian. We refer to the value of the maximal overlap as the ``GS fidelity''. As the Trotter number increases, the effective ground state at each QAOA step should converge to the ground state of the exact Hamiltonian, since the Trotterized mixer becomes more accurate. As demonstrated in Fig.~\ref{fig:unitary_gsfidelity_T}, even in the presence of potentially large Trotter error, a small Trotter number is sufficient for the effective ground state to be very close to the ground state of an exact mixing Hamiltonian. In other words, the GS fidelity converges much faster than the Trotter error with respect to the Trotter number. This observation is also reported in~\cite{yi2021success}.

\section*{Data availability}
Data for reproducing all the portfolio optimization results is available upon request from the authors.

\section*{Code availability}
The code for numerical simulations is available upon reasonable request.

\section*{Acknowledgments}
The authors thank Brian Neyenhuis, Jenni Strabley and the whole Quantinuum team for their support and feedback, and especially for providing us preview access to the Quantinuum H2-1 with 32 qubits. The authors thank their colleagues at the Global Technology Applied Research center of JPMorgan Chase for helpful discussions. Disclaimer: This paper was prepared for informational purposes by the Global Technology Applied Research center of JPMorgan Chase \& Co. This paper is not a product of the Research Department of JPMorgan Chase \& Co. or its affiliates. Neither JPMorgan Chase \& Co. nor any of its affiliates makes any explicit or implied representation or warranty and none of them accept any liability in connection with this paper, including, without limitation, with respect to the completeness, accuracy, or reliability of the information contained herein and the potential legal, compliance, tax, or accounting effects thereof. This document is not intended as investment research or investment advice, or as a recommendation, offer, or solicitation for the purchase or sale of any security, financial instrument, financial product or service, or to be used in any way for evaluating the merits of participating in any transaction.

\section*{Author Contributions}
R.S. conceived the idea. Z.H. and Y.S. developed the simulation code. Z.H. performed the numerical simulations. Z.H., C.L. and R.S. performed experiments on the trapped-ion quantum processor. All authors participated in technical discussions and contributed to the writing of the manuscript.

\section*{Competing Interests}
The authors declare no competing financial or non-financial interests.

\bibliography{reference} %

\end{document}